\UseRawInputEncoding
\documentclass[letterpaper, 10 pt, conference]{ieeeconf}  

\IEEEoverridecommandlockouts                              
\usepackage{cite}
\usepackage{amsmath,amssymb,amsfonts}
\usepackage{algorithmic}
\usepackage{graphicx}
\usepackage{threeparttable}
\usepackage{enumerate}
\usepackage{subfigure}
\usepackage{multirow}
\usepackage{textcomp}
\usepackage{xcolor}

\title{\LARGE \bf
Emotion-Conditioned Melody Harmonization with Hierarchical Variational Autoencoder
}

\author{Shulei Ji$^{*}$ and Xinyu Yang$^{*}$
\thanks{$^{*}$Shulei Ji and Xinyu Yang are with the School of Computer Science and Technology, Xi'an Jiaotong University, Xi'an, 710049, China (e-mail: 
        {\tt\small taylorji@stu.xjtu.edu.cn}; {\tt\small yxyphd@mail.xjtu.edu.cn})}%
\thanks{(Corresponding author: Xinyu Yang)}%
}

\begin{document}

\maketitle
\thispagestyle{empty}
\pagestyle{empty}

\begin{abstract}
Existing melody harmonization models have made great progress in improving the quality of generated harmonies, but most of them ignored the emotions beneath the music. Meanwhile, the variability of harmonies generated by previous methods is insufficient. To solve these problems, we propose a novel LSTM-based Hierarchical Variational Auto-Encoder (LHVAE) to investigate the influence of emotional conditions on melody harmonization, while improving the quality of generated harmonies and capturing the abundant variability of chord progressions. Specifically, LHVAE incorporates latent variables and emotional conditions at different levels (piece- and bar-level) to model the global and local music properties. Additionally, we introduce an attention-based melody context vector at each step to better learn the correspondence between melodies and harmonies. Objective experimental results show that our proposed model outperforms other LSTM-based models. Through subjective evaluation, we conclude that only altering the types of chords hardly changes the overall emotion of the music. The qualitative analysis demonstrates the ability of our model to generate variable harmonies.
\end{abstract}

\section{Introduction} \label{section1}
Melody harmonization is a conditional music generation task to generate chord progression conditioned on the given melody \cite{6-2,6-1}. This task aims at generating human-like harmonies to enrich monotonous melodies and enhance musical expression. In addition to improving computer creativity, understanding human composition mechanisms, and bringing inspiration to composers, melody harmonization can also be applied in areas such as entertainment and education. 

Many previous neural-network-based melody harmonization methods used recurrent neural networks \cite{9,10}, e.g., Long Short-Term Memory (LSTM), to generate the corresponding chords for the melody input, which hardly generated variable results for the same input and ignore the learning of chord transitions. Recently, the variational auto-encoder (VAE) \cite{11,12} has been utilized to generate coherent and varied harmonies. However, these studies all neglected the emotion conveyed by the harmonies.    

Generating music conditioned on emotions using deep learning techniques, also known as emotion-conditioned music generation, has gradually become a trend \cite{1,2,3}, where researchers tend to generate music from scratch based on the given emotions. Nevertheless, few studies explore the impact of emotion on conditional music generations, i.e., generating partial music given other musical contexts, such as melody harmonization, the focus of this paper. 

The role of emotional condition in the melody harmonization task warrants careful consideration. There are two cases: firstly, the harmony exhibits distinct emotions, so it maybe feasible to change the emotion of the overall lead sheet by generating harmony with specified emotion; secondly, in cases where discerning the emotion of the harmony is challenging, the emotional conditions can serve to assist in generating better harmonies. 

Our previous work \cite{0} focused on the second case, which proposed an emotion-conditioned hierarchical transformer VAE named EmoMusicTV to explore the impact of emotional conditions on multiple tasks on lead sheets. It has been proven that introducing emotions into the model can benefit the generated results for two conditional generation tasks, i.e., melody harmonization and melody generation given harmony. Note that the above practice is based on our empirical findings that melody seemingly dominates in conveying the emotion of music rather than harmony. However, this assumption has not been experimentally verified. 

In addition, the transformer-based EmoMusicTV was compared with two LSTM-based methods \cite{13,11} for the melody harmonization task in our previous work \cite{0}, one of which was also based on VAE generative model. However, it is difficult for such a comparison to highlight the advantages of the hierarchical hidden variable structure, since the transformer appears to be superior in sequence learning when compared to LSTM. Therefore, it is hard to say whether the advantages of EmoMusicTV come from the transformer or the hierarchical hidden variable structure.

To make up for the deficiency in previous work, a novel LSTM-based Hierarchical VAE (LHVAE) is proposed to generate harmonies with emotional conditions at different levels and improve the diversity of the generated harmonies. Following EmoMusicTV \cite{0}, LHVAE also incorporates a hierarchical latent variable structure, where a piece-level latent variable learns holistic properties and a chain of bar-level latent variables captures short-term variations. Emotional conditions at different levels are embedded in their corresponding latent spaces neatly. The comparison between LHVAE and other LSTM-based models can better highlight the effectiveness of hierarchical latent variable structure since all the models adopt the same backbone, i.e., LSTM.
\begin{figure*}
	\setlength{\abovecaptionskip}{0cm}
	\setlength{\belowcaptionskip}{0cm}
	\subfigure[Event-based music representation.]{
		\setlength{\abovecaptionskip}{0cm}
		\setlength{\belowcaptionskip}{0cm}
		\begin{minipage}[t]{0.3\linewidth}
			\setlength{\abovecaptionskip}{0cm}
			\setlength{\belowcaptionskip}{0cm}
			\centering
			\includegraphics[width = 1\linewidth]{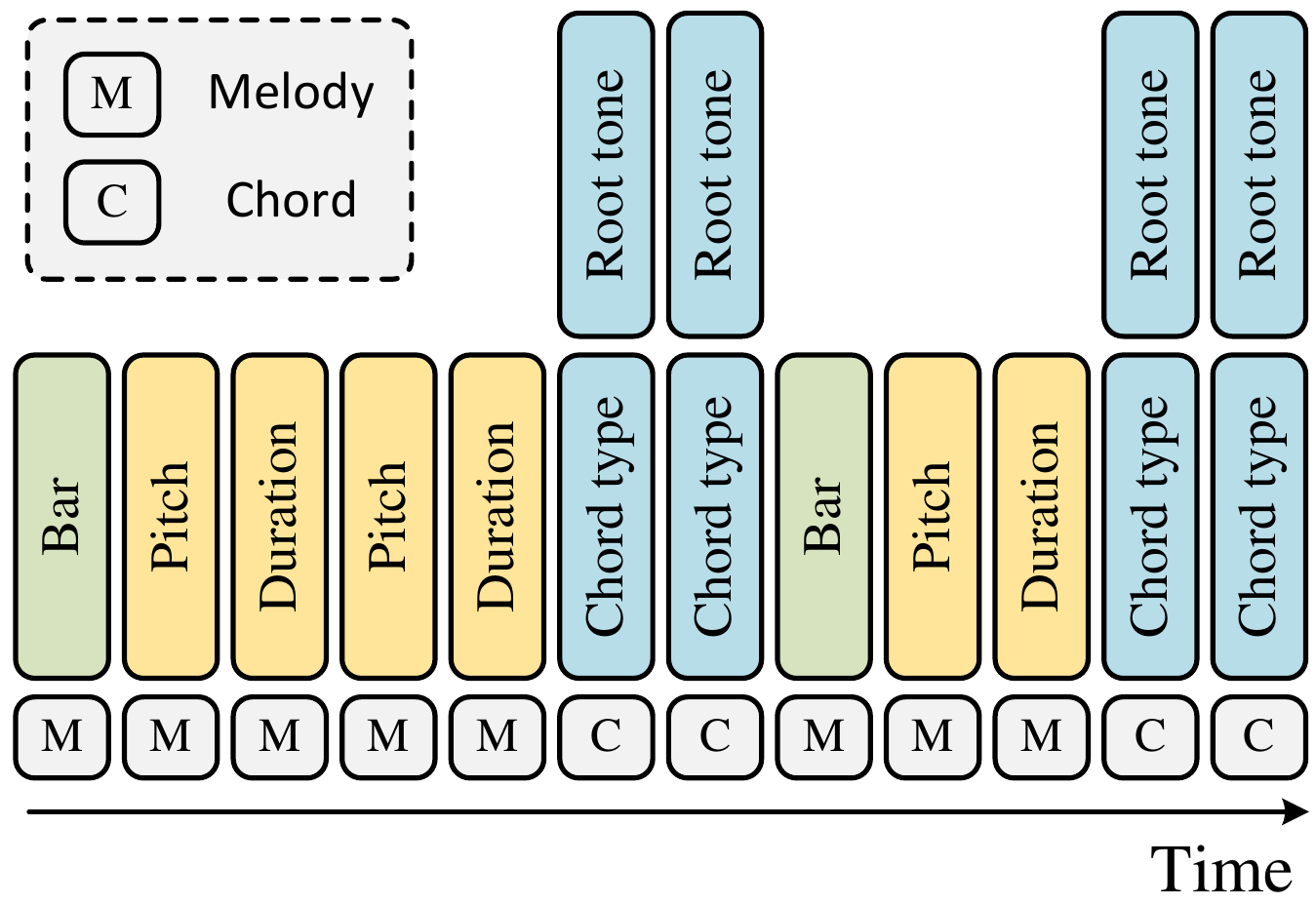}
			\label{fig1-1}
		\end{minipage}
	}
	\centering
	\subfigure[The architecture of the LSTM-based hierarchical VAE. The model first produces a piece-level latent variable, and then iteratively generates bar-level latent variables at the beginning of each bar.]{
		\setlength{\abovecaptionskip}{0cm}
		\setlength{\belowcaptionskip}{0cm}
		\begin{minipage}[t]{0.635\linewidth}
			\centering
			\includegraphics[width = 1\linewidth]{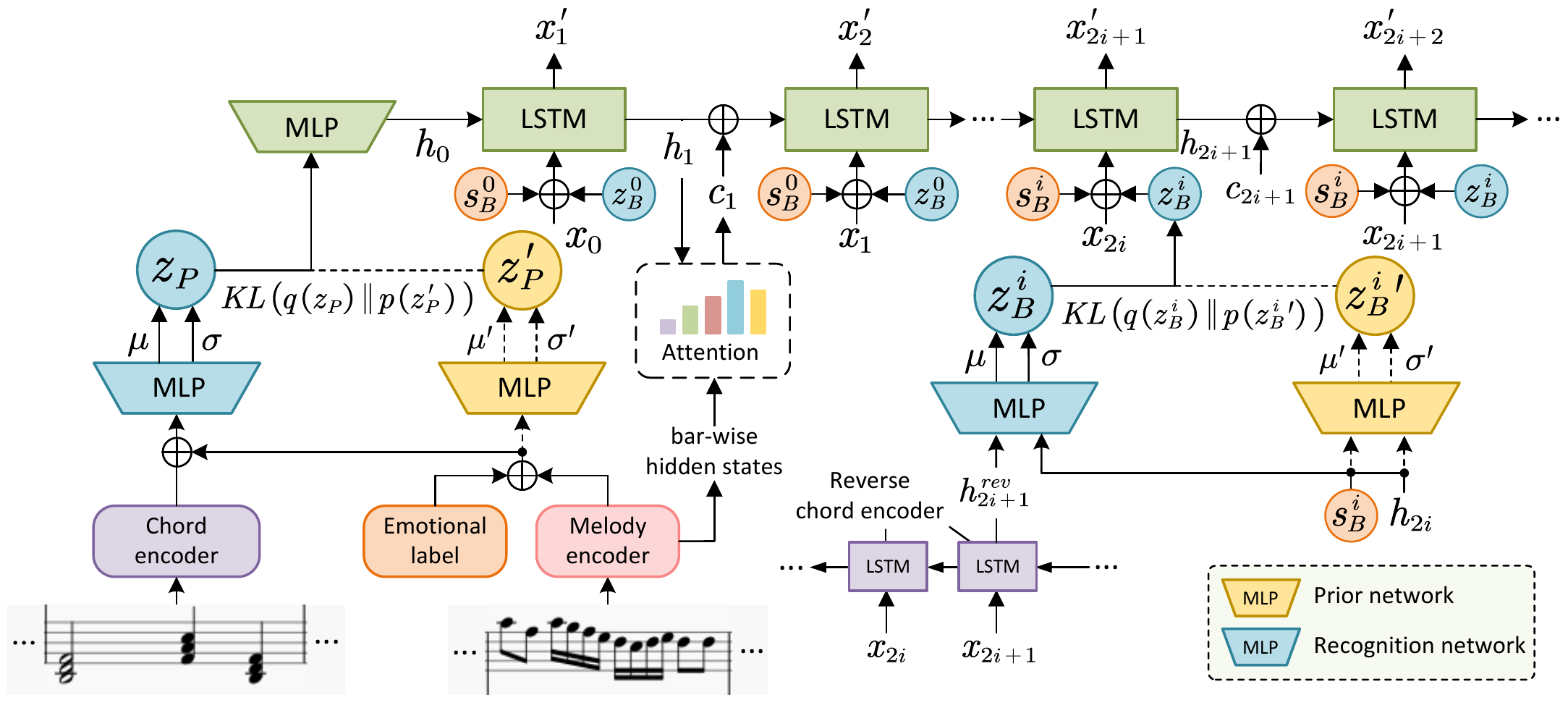}
			\label{fig1-2}
		\end{minipage}
	}
	\caption{The adopted music representation and the proposed model.}
	\label{fig1}
	\vspace{-4mm}
\end{figure*} 

We conduct a series of experiments to verify the effectiveness of our proposed model for melody harmonization and to answer the following question of concern: \textbf{Whether the harmonies generated under different emotional conditions for the same melody can change the overall emotion of music}. Experimental results show that our proposed model outperforms other LSTM-based methods via objective evaluation and brings music variability through qualitative analysis. The subjective evaluation shows that it is hard to change the musical emotion perceived by humans via only altering the chord progressions\footnote{Note that in this paper we only change the types of chords, regardless of the density of chords, i.e., each bar contain two chords.\label{fn1}}. 

\section{Related Work}  \label{section2}
Deep neural networks have been widely applied for automatic melody harmonization. Lim et al. \cite{9} utilized a two-layer bidirectional LSTM (BiLSTM) network to learn the correspondence between melody and chord sequences. Yeh et al. \cite{10} proposed an extended BiLSTM model to predict not only chord labels but also chord functions. Sun et al. \cite{13} input the melody and the partially masked chord sequence into the BiLSTM to learn the missing parts in the chord sequence. Then they adopted blocked Gibbs sampling for inference. Chen et al. \cite{11} proposed a VAE-based melody harmonization model conditioned on surprise contours. Rhyu et al. \cite{12} proposed the regularized variational Transformer to control the attribute of generated chords. Zeng et al. \cite{15} proposed a novel reinforcement learning method to first learn a refined structured melody representation, and then to generate chords for each learned segment representation. Our previous work \cite{-1} also proposed a novel melody harmonization system based on reinforcement learning, which contains three well-designed reward modules for introducing other guidance, such as music theory knowledge. However, none of the above methods explore melody harmonization conditioned on emotions. 

To the best of our knowledge, only Takahashi et al. \cite{16} and EmoMusicTV \cite{0} generated harmonic arrangements for the given melodies with emotional constraints. Takahashi et al. \cite{16} adopted crowd-sourced mood tags as emotion labels, which are considered as ``weak labels''. They found that the errors between the perceived and the target emotions remain relatively large and it is difficult to prove that the model consistently expresses the desired emotions. EmoMusicTV \cite{0} empirically assumed that melody dominates in conveying the emotion of music and took emotional conditions as auxiliary information to improve the quality of the generated harmonies. However, neither of the above researches addressed our proposed question, i.e., the possibility of changing the overall emotion of music via altering chord progressions\textsuperscript{\ref{fn1}}. 
\section{Methodology} \label{section3}
\subsection{Music Representation}
Same as \cite{0}, we adopt the event-based music representation \cite{6-2} to represent the lead sheets, as illustrated in Fig. \ref{fig1-1}. Each melody note is represented using a pair of consecutive tokens, namely the \textit{pitch} and \textit{duration} tokens. The beginning of a bar is indicated by the \textit{bar} token. Each chord is represented by a super token \cite{26} consisting of \textit{chord type} and \textit{root tone}. The numbers of \textit{pitch} (plus REST) and \textit{duration} events are 61 and 37, respectively. There are 7 distinct \textit{chord type} events (including REST) and 41 different \textit{root tone} events. Consequently, the melody event is encoded as a 99-dimensional one-hot vector, while the chord event is represented by a multi-hot vector of 48 dimensions.
\subsection{LSTM-based Hierarchical VAE}
We propose a novel LSTM-based Hierarchical VAE (LHVAE) for emotion-conditioned melody harmonization to improve the variability of the generated chord progressions and to explore the impact of emotional conditions on the generated harmonies, as shown in Fig. \ref{fig1-2}. The hierarchical latent variable structure in LHVAE follows the design of EmoMusicTV \cite{0}.

We represent the melody sequence as $Y=\{y_1,y_2,...,y_m\}$, the chord sequence as $X=\{x_1,x_2,...,x_n\}$, where $m$ and $n$ are the number of melody events and chords. The proposed model contains latent variables at different levels, i.e., piece-level latent variable $z_P$ and bar-level latent variables $Z_B=\{z_B^1,\ldots,z_B^I\}$, $I$ is the number of bars. The piece-level emotion $s_P$ and emotion expressed in each bar $S_B=\{s_B^1,\ldots,s_B^I\}$ are embedded in their corresponding latent spaces to guide the chord generation. We use $S$ to represent all the emotions in the follow-up.

The latent variables exhibit structured dependencies that each bar-level latent variable $z_B^i$ is conditioned not only on $z_P$ but also on the previous latent trajectories $Z_B^{\textless i}$. By variational inference, we derive the objective function which is the Evidence Lower Bound ($ELBO$) of the conditional log likelihood of $p(X|Y,S)$
\begin{equation}
\label{eq2}
\begin{split}
\log p(X|Y,S)&\geq \textstyle\sum_i\mathbb{E}_{q_\phi(Z_B^{\textless i}|X,z_P,s_P,S_B,Y)}[{ELBO_i}]-\\ 
&\quad\ KL[q_\phi(z_P|s_P,X,Y)\|p_\varphi(z_P|s_P,Y)]\\ 
&=-\mathcal{L}(X;Y,S,\theta,\phi,\varphi)  
\end{split}
\end{equation}
where KL is the Kullback-Leibler divergence. The loss $\mathcal{L}(X;Y,S,\theta,\phi,\varphi)$ consists of a piece-level KL term and the sum of negative bar-level $ELBO_i$. The $ELBO_i$ can be factorized as:
\begin{equation}
\label{eq3}
\resizebox{.9\hsize}{!}{$
\begin{split}
ELBO_i=&\mathbb{E}_{q_\phi(z_B^i|X,U)}[\log P_\theta(X^i|X^{\textless i},z_B^i,U)]-KL_i\\ 
KL_i=&KL[q_\phi(z_B^i|X,U)\|p_\varphi(z_B^i|X^{\textless i},U)]
\end{split}
$}
\end{equation}
where $U$ refers to the union of $Y, z_P,s_P,Z_B^{\textless i},$ and $S_B^{\leq i}$, $X^i$ and $X^{\textless i}$ represent the chords within $i$th bar and the chords before $i$th bar, respectively. $\log P_\theta$ is the log-likelihood of the predicted probabilities, and $p_\varphi$ and $q_\phi$ are the prior distribution and the approximate posterior distribution, respectively.

We assume that the latent variables at different levels obey the isotropic Gaussian distribution, like previous work \cite{34} did. The mean $\mu$ and standard deviation $\sigma$ are generated by the Multi-Layer Perceptron (MLP)-based prior model and recognition model \cite{35}, as depicted in Fig. \ref{fig1-2}. The piece-level $z_P$ is computed only once at the beginning of generating the entire harmony, which initializes the first hidden state $h_0$ of the LSTM. While the bar-level latent variables $Z_B$ are computed at the beginning of each bar. Note that each bar contains two chords in this paper.

Referring to Eq. \eqref{eq2}, the piece-level prior model $p_\varphi$ takes as input the melody and the piece-level emotion to obtain $z_P$, and the piece-level recognition model $q_\phi$ additionally needs the chord. According to the Eq. \eqref{eq3}, the bar-level prior model $p_\varphi$ takes historical output $X^{\textless i}$ and other elements as input to get $z_B^i$, while the bar-level recognition model $q_\phi$ requires the entire output sequence $X$. Thus, we add a reverse chord encoder to model output sequences $X^{\geqslant i}$ that have not been generated yet. For code implementation, the last hidden state $h_{2i}$ of LSTM, implicitly containing $X^{\textless i}, Y, z_P,s_P,Z_B^{\textless i},$ and $S_B^{\textless i}$, are input into the bar-level prior and recognition network, as illustrated in Fig. \ref{fig1-2}. 
\begin{figure}[t]
	\setlength{\abovecaptionskip}{0cm}
	\setlength{\belowcaptionskip}{0cm}
	\centering
	\includegraphics[width=1\linewidth]{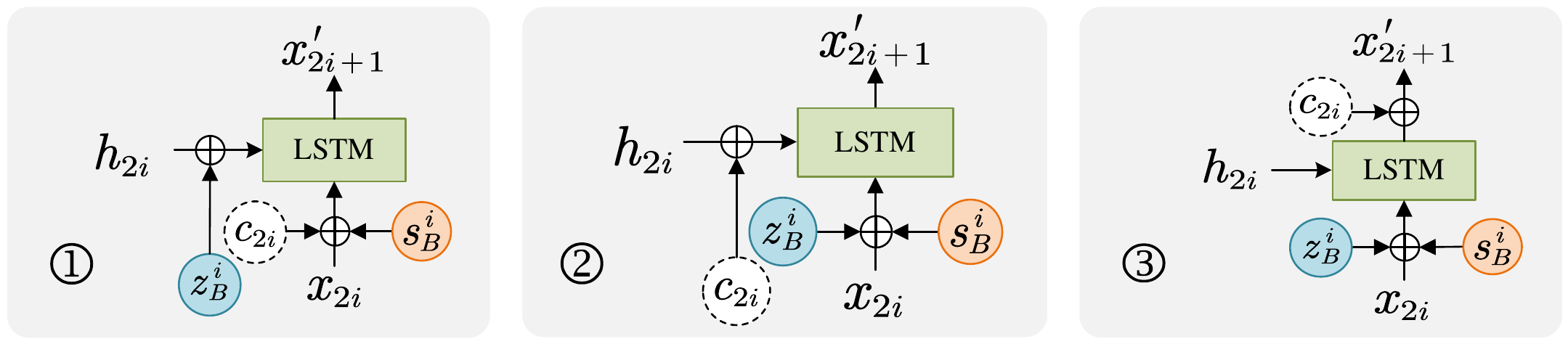}
	\caption{Three ways to inform the model bar-level hidden variable and melody context vector. $z_B^i$ and $s_B^i$ are latent variable and emotional label of the $i$th bar, respectively. $c_{2i}$ is the melody context vector at time step $2i$.}
	\label{fig2}
	\vspace{-3mm} 
\end{figure}

Additionally, to better learn the correspondence between melody and chord, we calculate the melody context vector $c_t$ at each step by making the hidden state $h$ of LSTM account for the melody hidden states $h^Y$ within the current bar. 
\begin{equation}
c_t=\sum_{j=m^i_{start}}^{m^i_{end}}\alpha_{tj}h_j^Y,\ \alpha_{tj}=f_{attention}(h_{t-1},h_j^Y)
\end{equation}
where $h_j^Y$ is the $j$th hidden states of melody encoder, $m^i_{start}$ and $m^i_{end}$ indicate start and end of melody sequences within the $i$th bar. $f_{attention}$ is a function that performs $softmax$ after the inner product, and $\alpha_{tj}$ is the attention weight. 

At each step, we feed the bar-level latent variable, the context vector and the emotional label together into the model to generate chords. As shown in Fig. \ref{fig2}, we provide three ways to inform the model bar-level latent variable $z_B$ and melody context vector $c$. The experimental results will prove that the second one is the best choice, so the model diagram in Fig. \ref{fig1-2} is drawn according to the second type.
\begin{table}[t]
	\setlength{\abovecaptionskip}{0cm}
	\setlength{\belowcaptionskip}{0cm}
	\caption{Holistic emotion distribution of NMD dataset.}
	\begin{minipage}[l]{1\linewidth}
		\centering
		\renewcommand{\arraystretch}{1.2}
		\resizebox{0.75\textwidth}{!}{%
			\begin{tabular}{|c|c|c|}
				\hline
				Negative & Neutral & Positive \\ \hline
				624 (5.52\%)     &1272 (11.25\%)    &9408 (83.23\%) \\ \hline 
			\end{tabular}%
		}
		\label{tab1}
	\end{minipage}      
	\vspace{-3mm}
\end{table}
\begin{table*}[t]
	\setlength{\abovecaptionskip}{0cm}
	\setlength{\belowcaptionskip}{0cm}
	\caption{The results of three ways informing model latent variables and context vectors. The metric values closest to GT are marked in \textbf{bold}. We adopt two-tailed t-test scores to determine whether there is a significant difference between the objective results of the model and GT. The objective results of the models are expected to be not significantly different from those of GT, i.e., \textit{p} $\geqslant0.05$. All p-values equal to or greater than 0.05 are marked in \textbf{bold}.}
	\begin{minipage}[l]{1\linewidth}
		\centering
		\renewcommand{\arraystretch}{1.2}
		\resizebox{1.0\textwidth}{!}{%
			\begin{tabular}{l|r|r|r|r|r|r|r|r|r|r|r|r}
				\hline
				Models      & CHE    &p-value         & CC    &p-value         & CTD      &p-value       & CTnCTR     &p-value     & PCS        &p-value     & MCTD      &p-value    \\ \hline
				GT          & 0.7493$\pm$0.20    &-      & 6.3120$\pm$2.04     &-      & 0.1034$\pm$0.07      &-    & 0.4214$\pm$0.11       &-    & 0.1801$\pm$0.40       &-    & 0.0999$\pm$0.03     &-    \\ \cline{1-13} 
				LHVAE1     &\textbf{0.7083$\pm$0.20}    &1.2e-05        &\textbf{7.9930$\pm$2.78}  &2.9e-31  &0.0257$\pm$0.04  &6.7e-56  &\textbf{0.4214$\pm$0.11}  &\textbf{0.1670}  &0.1687$\pm$0.37   &0.0032 &0.1007$\pm$0.03  &1.2e-36 \\ \cline{1-13} 
				LHVAE2     &0.7993$\pm$0.22      &4.5e-21     &8.8050$\pm$3.09 &1.1e-47 &\textbf{0.0912$\pm$0.05}  &\textbf{0.3107} &0.4212$\pm$0.11 &\textbf{0.2322} &\textbf{0.1872$\pm$0.41} &\textbf{0.0530} & \textbf{0.0995$\pm$0.03}  &1.2e-14 \\ \cline{1-13} 
				LHVAE3 &0.8122$\pm$0.23     &3.1e-25      & 8.9900$\pm$2.68    &2.4e-61       &0.0788$\pm$0.04    &2.5e-07      & \textbf{0.4214$\pm$0.11}      &\textbf{0.6367}     & 0.1905$\pm$0.42    &0.0415       & 0.0992$\pm$0.03    &6.2e-19 \\ \hline 
			\end{tabular}%
		}
		\label{tab2}
	\end{minipage}      
\end{table*} 
\begin{table*}[t]
	\setlength{\abovecaptionskip}{0cm}
	\setlength{\belowcaptionskip}{0cm}
	\caption{Model comparison results.}
	\begin{minipage}[l]{1\linewidth}
		\centering
		\renewcommand{\arraystretch}{1.2}
		\resizebox{1.0\textwidth}{!}{%
			\begin{threeparttable}
				\begin{tabular}{l|r|r|r|r|r|r|r|r|r|r|r|r}
					\hline
					Models      & CHE    &p-value         & CC    &p-value         & CTD      &p-value       & CTnCTR     &p-value     & PCS        &p-value     & MCTD      &p-value    \\ \hline
					GT          & 0.7493$\pm$0.20    &-      & 6.3120$\pm$2.04     &-      & 0.1034$\pm$0.07      &-    & 0.4214$\pm$0.11       &-    & 0.1801$\pm$0.40       &-    & 0.0999$\pm$0.03     &-    \\ \cline{1-13} 
					BGS\cite{13}\tnote{*}     &0.6994$\pm$0.17    &3.2e-10        &\textbf{5.5820$\pm$1.77}  &\textbf{0.0746}  &0.1407$\pm$0.06  &3.8e-28  &0.4216$\pm$0.11  &\textbf{0.0879}  &0.1875$\pm$0.40   &0.0038 &0.0994$\pm$0.03  &2.1e-27 \\ \cline{1-13} 
					SurpriseNet\cite{11}\tnote{*} &\textbf{0.7964$\pm$0.21}     &1.8e-10      & 7.9980$\pm$1.79    &2.6e-06       &0.1193$\pm$0.05    &5.7e-07       & 0.4212$\pm$0.11      &\textbf{0.1717}     & \textbf{0.1834$\pm$0.41}    &\textbf{0.1238}       & \textbf{0.0997$\pm$0.03}    &8.3e-05 \\ \cline{1-13} 
					LHVAE2     &0.7993$\pm$0.22      &1.0e-12     &8.8050$\pm$3.09 &1.1e-47 &\textbf{0.0912$\pm$0.05}  &\textbf{0.3107} &\textbf{0.4212$\pm$0.11} &\textbf{0.2322} &0.1872$\pm$0.41 &\textbf{0.0530} & 0.0995$\pm$0.03  &1.2e-14 \\ \hline 
				\end{tabular}%
				\begin{tablenotes}
					\footnotesize
					\item[*] The results of BGS\cite{13} and SurpriseNet\cite{11} are derived from \cite{0}.
				\end{tablenotes}
			\end{threeparttable}
		}
		\label{tab3}
	\end{minipage}      
	\vspace{-3mm}
\end{table*}
\vspace{-3mm}
\subsection{Loss function} \label{3-C}
The output of LSTM at each step will be fed into two feed-forward heads to predict \textit{chord type} and \textit{root tone}, respectively. To alleviate the unbalanced emotion distribution caused by the uneven number of chord types (see Table \ref{tab1}), we use the chord balancing trick proposed by Sun et al. \cite{13}. Specifically, we assign a weight $w_c$ for each \textit{chord type} $c$ in the negative log-likelihood (NLL) loss of chord type.
\begin{equation}
\begin{split}
L\left( X;Y,S,\theta \right)&=-\sum_c{w_cx_{t,c}\log P_{\theta}\left( x_{t,c}|x_{<t},Y,S \right)} 
\\
w_c&=\left| C \right|\times \frac{{1}/{\left( 1e5+n_c \right)}}{\sum_{c'}{1}/{\left( 1e5+n_{c'} \right)}}
\end{split}
\end{equation}
where $x_{t,c}$ is the $c$-th element of the one-hot vector $x_t$ of chord type and $\theta$ represents the model parameter. $\left| C \right|$ is the number of chord types (i.e., 7), $n_c$ is the number of chord type $c$ in the training data and $1e5$ is a temperature-like parameter for count smoothing. Note that the reconstruction loss for the \textit{root tone} is the unweighted NLL loss.

\section{Experiments} \label{section4}
\subsection{Experimental Setup}
\subsubsection{Dataset}
We adopt the preprocessed Nottingham Music Dataset\footnote{abc.sourceforge.net/NMD/} (NMD) by \cite{0} with the annotated piece- and bar-level emotional labels. One difference is that we simplify five emotion categories (very negative, moderate negative, neutral, moderate positive, and very positive) in \cite{0} into three categories: negative (0), neutral (1), and positive (2). The holistic emotion distribution of the dataset is presented in Table \ref{tab1}, from which we see that the number of musical pieces with positive emotion is much larger than that of neutral and negative emotions, and the number of negative pieces is the least. Since the emotional labels are computed from chords using the method of Makris et al.\cite{5}, the unbalanced distribution of emotions is caused by the uneven number of different chord types. Thus, we used chord balancing \cite{3} to alleviate this problem in this paper (Section \ref{3-C}). Note that each bar contains two chords. 
\subsubsection{Implementation Details}
The encoders and decoder in the LHVAE model are all 2-layer LSTMs. The size of the hidden state, latent space and batch are 256, 128 and 24, respectively. The learning rate is 3e-4 and the model is trained with the Adam optimizer. We use the KL annealing trick \cite{30,32} to alleviate the posterior collapse problem \cite{27} in VAE training. Specifically, the piece- and bar-level KL weights start from zero and linearly increase 1e-5 and 1e-4 per epoch, respectively. The dataset is randomly divided into 90\% training set and 10\% validation set for objective evaluation. The emotional labels in the evaluation stage are derived from the validation set.
\subsubsection{Compared Models}
We compare our model with Ground-Truth (GT) and other two LSTM-based melody harmonization models.


\textbf{BGS}\cite{13}: A BiLSTM model with orderless NADE training process and chord balancing, then using blocked Gibbs sampling (BGS) for inference.

\textbf{SurpriseNet}\cite{11}: A BiLSTM-based CVAE taking surprise contours as conditions to generate chord sequences. 
\subsubsection{Evaluation Metrics}
To evaluate the generated chords quantitatively, we use six music metrics defined by Yeh et al. \cite{10}, i.e., Chord histogram entropy (CHE), Chord coverage (CC), Chord tonal distance (CTD), Chord tone to non-chord tone ratio (CTnCTR), Pitch consonance score (PCS) and Melody-chord tonal distance (MCTD). Among them, CC, CHE and CTD evaluate chords in terms of the number of chord types, chord distribution and chord transition, while the rest metrics evaluate the consistency and harmonicity between melody and harmony. Note that the generated results are generally better when the metrics values are closer to the GT. Besides, we use four criteria \cite{13} to evaluate the generated chords subjectively, i.e., Chord progression, Coherence, Interestingness, and Overall.
%
%
%
\begin{table}[t]
	\setlength{\abovecaptionskip}{0cm}
	\setlength{\belowcaptionskip}{0cm}
	\centering
	\caption{Emotional control accuracy.}
	\begin{minipage}[l]{0.7\linewidth}
		\renewcommand{\arraystretch}{1.2}
		\resizebox{1.0\textwidth}{!}{%
			\begin{tabular}{l|r|r|r}
				\hline
				Accuracy      & LHVAE1    & LHVAE2         & LHVAE3   \\ \hline
				Piece-level   & \textbf{0.98}    & 0.95      & 0.89  \\ \cline{1-4} 
				Bar-level     & \textbf{0.96}    &0.93       &0.88 \\ 
				\hline
			\end{tabular}%
		}
		\label{fig3}
	\end{minipage}   
	\vspace{-3mm}
\end{table}
\begin{table*}[t]
	\setlength{\abovecaptionskip}{0cm}
	\setlength{\belowcaptionskip}{0cm}
	\caption{Ablation study results of emotional conditions and latent variables. $S$ denotes the emotional conditions. $z_P$ is the piece-level latent variable, $Z_B$ is the chain of bar-level latent variables, and $Z_A$ represents all the latent variables.}
	\begin{minipage}[l]{1\linewidth}
		\centering
		\renewcommand{\arraystretch}{1.2}
		\resizebox{.95\textwidth}{!}{%
			\begin{tabular}{l|r|r|r|r|r|r|r|r|r|r|r|r}
				\hline
				Models      & CHE    &p-value         & CC    &p-value         & CTD      &p-value       & CTnCTR     &p-value     & PCS        &p-value     & MCTD      &p-value    \\ \hline
				GT          & 0.7493$\pm$0.20    &-      & 6.3120$\pm$2.04     &-      & 0.1034$\pm$0.07      &-    & 0.4214$\pm$0.11       &-    & 0.1801$\pm$0.40       &-    & 0.0999$\pm$0.03     &-    \\ \cline{1-13} 
				LHVAE2     &0.7993$\pm$0.22      &1.0e-12     &8.8050$\pm$3.09 &1.1e-47 &\textbf{0.0912$\pm$0.05}  &\textbf{0.3107} &0.4212$\pm$0.11 &\textbf{0.2322} &0.1872$\pm$0.41 &\textbf{0.0530} & 0.0995$\pm$0.03  &1.2e-14 \\ \cline{1-13} 
				w/o $S$ &0.7395$\pm$0.21     &\textbf{0.0637}      & 7.8660$\pm$2.99    &1.6e-18       &0.0522$\pm$0.06    &5.9e-21  & 0.4211$\pm$0.11      &0.0392     & 0.1704$\pm$0.38    &0.0089       & 0.1006$\pm$0.03    &2.8e-24 \\ \cline{1-13} 
				w/o $z_P$    &0.7746$\pm$0.21      &1.9e-12     &8.6660$\pm$3.13 &1.5e-44 &0.0836$\pm$0.05  &0.0343 &0.4213$\pm$0.11 &\textbf{0.4765} &0.1880$\pm$0.40 &0.0047 &\textbf{0.0999$\pm$0.03}  &\textbf{0.0657} \\ \cline{1-13}
				w/o $Z_B$    &0.7562$\pm$0.20      &1.8e-10     &7.0400$\pm$2.13 &2.6e-27 &0.0782$\pm$0.05  &0.0001 &0.4215$\pm$0.11 &0.0326 &\textbf{0.1774$\pm$0.40} &\textbf{0.6148} & 0.1000$\pm$0.03  &\textbf{0.9385} \\ \cline{1-13}
				w/o $Z_A$    &\textbf{0.7445$\pm$0.20}      &0.0008     &\textbf{6.6810$\pm$2.08} &1.9e-15 &0.0586$\pm$0.05  &1.4e-24 &\textbf{0.4214$\pm$0.11} &0.0201 &0.1756$\pm$0.39 &\textbf{0.9358} & 0.1001$\pm$0.03  &3.8e-05 \\
				\hline
			\end{tabular}%
		}
		\label{tab4}
	\end{minipage}      
	\vspace{-3mm}
\end{table*}
\subsection{Experimental Results and Analyses}
\begin{figure*}
	\setlength{\abovecaptionskip}{0cm}
	\setlength{\belowcaptionskip}{0cm}
	\centering
	\includegraphics[width=0.95\linewidth]{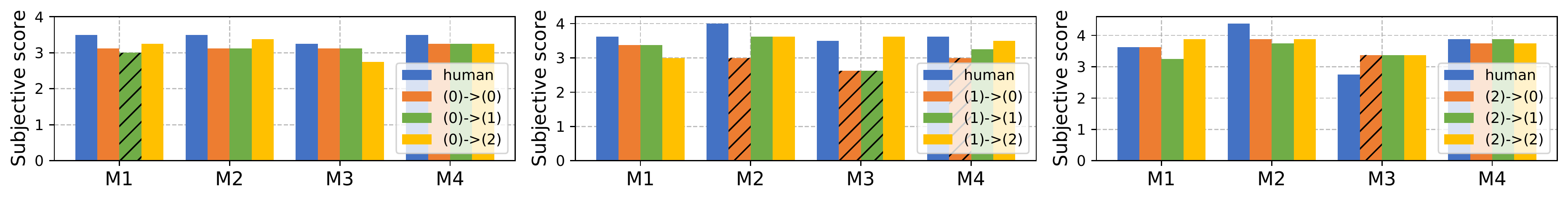}
	\caption{Histogram of human rating results. M1, M2, M3 and M4 represent chord progression, coherence, interestingness and overall, respectively. ($s_{GT}$)$\rightarrow$($s_{Given}$) in the legend denotes music with ground-truth annotated emotion $s_{GT}$ is re-harmonized with given emotional conditions $s_{Given}$. The shadow coverage indicates significant differences from human-composed chords.}
	\label{fig5}       
	\vspace{-3mm}
\end{figure*}
\begin{figure}
	\setlength{\abovecaptionskip}{0cm}
	\setlength{\belowcaptionskip}{0cm}
	\centering
	\includegraphics[width=0.95\linewidth]{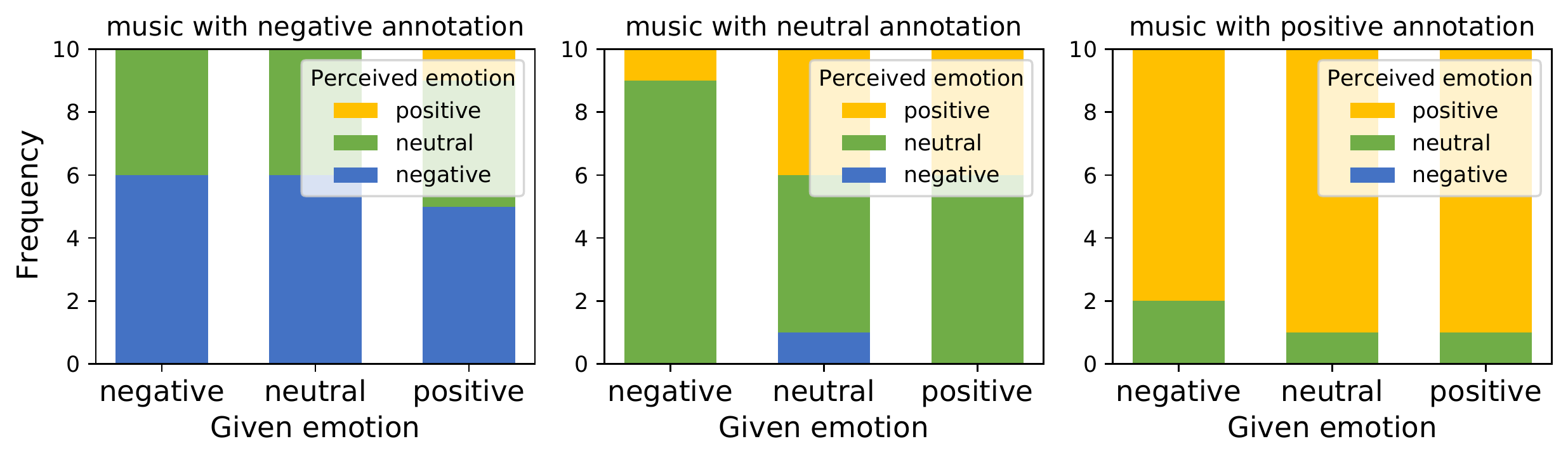}
	\caption{Emotion recognition results given by humans. The x-axis represents the given emotional labels, and the y-axis denotes the frequency of emotions perceived by the subjects.}
	\label{fig4}       
	\vspace{-3mm}
\end{figure}
\subsubsection{Objective Evaluation}
\paragraph{Comparative study}
We first investigate three ways (shown in Fig. \ref{fig2}) of feeding bar-level latent variable and melody context vector into the model, and the results are demonstrated in Table \ref{tab2}. We conclude that LHVAE2 gets the best results, especially on the metric evaluating chord transition (CTD). LHVAE1 perform better on CHE and CC, indicating better learning of chord distribution and the number of chord type. It seems that the concatenation of input and latent variables leads to better results of the melody-chord harmonicity (PCS, MCTD), while concatenating input with context vector leads to better CHE and CC. 

We also explored the emotional control capabilities of these three ways. Specifically, we generate harmonies for the melodies in the validation set given the annotated emotional labels. Then we calculate the piece- and bar-level emotions of the generated harmonies using the method proposed in \cite{5}. The emotional control accuracy is the number of generated harmonies conveying given emotions divided by the total number of pieces or bars, as shown in Table \ref{fig3}. LHVAE1 achieves the highest accuracy in both piece- and bar-level emotion control followed by LHVAE2. 

After making a trade-off between the objective results and the emotional control accuracy, we select LHVAE2 as the final version of the model. Then we compare it with other two LSTM-based models, as shown in Table \ref{tab3}. We can see that LHVAE2 excels at learning the chord transitions, showing the best CTD. And LHVAE2 is on par with SurpriseNet on learning the harmonicity between melody and chord (CTnCTR, PCS, MCTD). However, LHVAE2 and SurpriseNet perform poorly on CHE and CC, presenting higher values than BGS, which has no latent variable. We conjecture that more latent variables enable the generated chords more diverse than GT. 

It is worth noting that although LHVAE2 outperforms the other two LSTM models, it is inferior to the EmoMusicTV (compared with the 5th row in Table IV(a) in \cite{0}). This result shows that the advantages of EmoMusicTV come from the joint effect of the transformer and hierarchical hidden variable structure, and combining a stronger backbone with this structure can achieve better results.

\paragraph{Ablation study}
We first conduct an ablation study on LHVAE2 by eliminating the emotions, the results are shown in Table \ref{tab4}. We observe that removing emotions (w/o $S$) makes the results on CHE and CC improved, but the results of other metrics become worse. The conclusion comes that emotional conditions assist in improving the quality of generated chords in terms of chord transition and melody-chord harmonicity. This conclusion is slightly different from the results of EmoMusicTV w/o $S$ on the TheoryTab Dataset\footnote{https://www.hooktheory.com/theorytab} (TTD), which perform worse on almost all metrics (refer to 4th row in Table V(a) in \cite{0}). However, both experimental results demonstrate that emotional conditions can improve the generated harmonies more or less.

Additionally, we also show the effects of hierarchical latent variable structure in Table \ref{tab4}. Referring to the p-value, the lack of piece-level latent variables (w/o $z_P$) damages CTD and PCS, and removing bar-level latent variables (w/o $Z_B$ or $Z_A$) enables the results of CTD and CTnCTR far away from the GT. However, eliminating latent variables makes CHE, CC and MCTD closer to the GT than the complete LHVAE2. By and large, the combination of latent variables at different levels achieves better performance in learning chord transitions (CTD) and melody-chord harmonicity (CTnCTR and PCS), presenting insignificant differences with GT.
\subsubsection{Subjective Evaluation}
To answer the question introduced in Section \ref{section1} (i.e., whether the chord progressions generated under different emotional conditions for the same melody can change the overall emotion of music) and evaluate the generated chords subjectively, we invite human subjects to participate in two listening tests. In the first test, subjects are asked to identify the emotions for 9 pieces of music, which are generated by first selecting three lead sheets whose ground-truth holistic emotions are negative, neutral and positive respectively, then generating three harmonies for each melody given three different emotional conditions. Specifically, the piece- and bar-level emotions of the three emotional conditions are set to the same value at a time (i.e., all 0, all 1 or all 2)\footnote{The accuracy of bar-level positive, neutral and negative emotions for the generated chords are 97.98\%, 100\% and 88.89\%, respectively, which is computed as the number of generated emotions that are consistent with given emotions divided by the total number.}. For the second test, subjects are asked to rate 12 chord progressions according to the subjective criteria, including the above-mentioned nine and their ground-truth. A total of 20 subjects containing 10 males and 10 females participated in these two surveys. Their ages range from 18 to 30 years old, and six of them have learning experiences in musical instruments (e.g., piano and Chinese zither) and are familiar with harmonic theory more or less. 

The human emotion recognition results are shown in Fig. \ref{fig4}. The three sub-figures correspond to three melodies with negative, neutral and positive ground-truth emotions. We observe that although the melody is harmonized under different emotions, the music is still identified by the majority as its ground-truth emotion. It seems hard to change the overall emotion of music only by altering the simple pillar chords. On another front, the ground-truth emotional labels annotated using the method of \cite{5} from the harmony align with human annotation to some extent. 

The human rating results for the generated chord progressions are shown in Fig. \ref{fig5}. The three sub-figures still correspond to the three melodies with negative, neutral and positive ground-truth emotions. On the whole, human-composed harmonies are slightly better than the generated harmonies, and the quality of chords generated under different emotions has no obvious regularity. Through the two-tailed t-test, we find that in most cases, there are no significant differences between the generated and human-composed harmonies, and cases with significant differences are shaded in Fig. \ref{fig5}.
\begin{figure}
	\setlength{\abovecaptionskip}{0cm}
	\setlength{\belowcaptionskip}{0cm}
	\centering
	\hspace{-5mm}
	\includegraphics[width=0.8\linewidth]{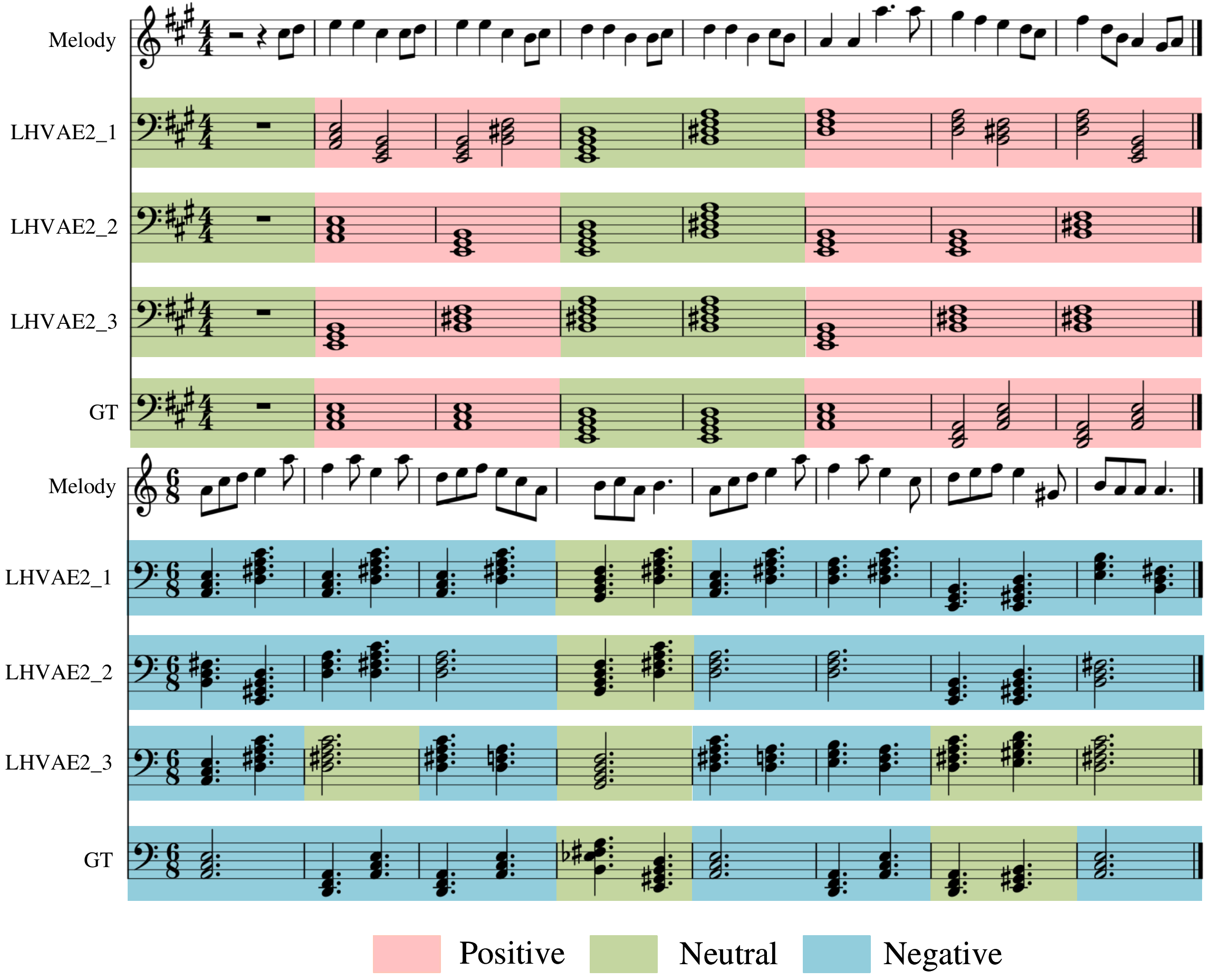}
	\caption{The visualizations of the generated chord progressions. For each piece of music, we show the original melody (top), the original harmony (bottom), and the harmonies generated by LHVAE2 (middle three).}
	\label{figure:6}       
	\vspace{-3mm}
\end{figure}
\subsubsection{Qualitative Analysis}
Given the inherent nature of VAE to sample from the latent spaces and introduce variability, it becomes possible to perform latent variable samplings multiple times and re-harmonize a given melody. Especially, the chain of bar-level latent variables in our proposed model makes the generated results more varied. Fig. \ref{figure:6} shows two pieces with chord progressions created by humans and the LHVAE2 model. It can be seen that our model generates chord progressions corresponding well to the ground-truth emotional labels and generates varied harmonies for the same melody while maintaining similar harmonic patterns. 
\section{Conclusion and Future Work}
In this paper, we make up for the deficiency of prior work on melody harmonization by proposing a novel LSTM-based VAE (LHVAE) model with a hierarchical latent variable structure and time-varying emotional conditions. We combine latent variables and emotions at different levels to produce harmonious, coherent, and diverse chord progressions. Experimental results show that LHVAE outperforms other LSTM-based melody harmonization models. We conclude that the guidance of emotional conditions can improve the quality of generated harmonies in terms of chord transition and harmonicity, while only altering chords hardly changes the overall emotion of music. For future work, we will explore the power of transformer-based hierarchical VAE and investigate music elements that have an impact on emotion (such as tempo, note density) to realize emotion-controllable music generation.


\bibliographystyle{IEEEtr} 
\bibliography{root_final}

\end{document}